\documentclass{article}

\RequirePackage[OT1]{fontenc}

\RequirePackage[lnms]{imsart}

\startlocaldefs
\def\A{{\cal A}}
\def\B{{\cal B}}

\def\D{{\cal D}}
\def\H{{\cal H}}

\def\S{{\cal S}}
\def\Th{\Theta}
\def\O{\Omega}
\def\d{\delta}
\def\l{\lambda}
\def\o{\omega}
\def\r{\rho}
\def\th{\theta}

\font\dik=msbm10

\def\CC{\hbox{{\dik C}}}
\def\NN{\hbox{{\dik N}}}
\def\PP{\hbox{{\dik P}}}
\def\EE{\hbox{{\dik E}}}

\def\ten{\otimes}
\def\one{{\bf 1}}
\def\id{{\rm id}}
\def\tr{{\rm tr}}

\def\bar{\overline}
\def\inp#1#2{\langle#1,#2\rangle}

\def\pure#1{|#1\rangle\langle#1|}
\def\tuple#1#2{#1_1,\ldots,#1_{#2}}
\def\li#1{\lim_{#1\to\infty}}
\def\sok#1{\sum_{#1=1}^k}
\def\sod#1{\sum_{#1=1}^d}
\def\Set#1#2{\left\{\;#1\;\bigg|\;#2\;\right\}}
\def\comprod#1#2#3{#1_{#2_{#3}}\circ\cdots\circ#1_{#2_1}}
\def\set#1#2{\{#1\,|\,#2\}}
\def\detpos{{\rm det}_{\rm pos}}

\def\Lin{\Lambda_{i_1,\ldots,i_n}}

\def\Pthn{\PP_{\theta_0}}
\def\square{{\vcenter{\vbox{\hrule height.4pt \hbox{\vrule width.4pt
height1.45ex \kern1.45ex \vrule width.4pt}
\hrule height.4pt}}}}
\def\qed{\hfill$\square$}
\newtheorem{theorem}{Theorem}
\newtheorem{lemma}[theorem]{Lemma}
\newtheorem{corollary}[theorem]{Corollary}
\newtheorem{definition}{Definition}

\endlocaldefs

\begin{document}

\begin{frontmatter}

\title{Purification of quantum trajectories}
\runtitle{Quantum trajectories}

\author{\fnms{Hans} \snm{Maassen}
\corref{Maassen}}\ead[label=e1]{maassen@math.ru.nl}
\thankstext{}{This paper is dedicated to Mike Keane on his 65-th birthday}
\address{Radboud University Nijmegen\\
\printead{e1}}
\author{\fnms{Burkhard} \snm{K\"ummerer}\ead[label=e2]{kuemmerer@mathematik.tu-darmstadt.de}}
\address{Technische Universit\"at Darmstadt\\
\printead{e2}}


\runauthor{Hans Maassen}

\begin{abstract}
We prove that the quantum trajectory of
repeated perfect measurement on a finite quantum system
either asymptotically purifies,
or hits upon a family of `dark' subspaces,
where the time evolution is unitary. 
\end{abstract}



\end{frontmatter}

\begin{section}
{Introduction}
A key concept in the modern theory of open quantum systems is the notion of
indirect measurement as introduced by Kraus \cite{Kraus}.
An {\em indirect measurement} on a quantum system is a (direct)
measurement of some quantity in its environment,
made after some interaction with the system has taken place.

\noindent
When we make such a measurement,
our description of the quantum system changes in two ways:
we account for the flow of time
by a unitary transformation (following Schr\"odinger),
and we update our knowledge of the system by conditioning
on the measurement outcome (following von Neumann).
If we then repeat the indirect measurement indefinitely,
we obtain a chain of random outcomes.
In the course of time we may keep record of the updated density
matrix $\Theta_t$,
which at time $t$ reflects our best estimate of all
observable quantities of the quantum system,
given the observations made up to that time.
This information can in its turn be used to predict later
measurements outcomes.
The stochastic process $\Theta_t$ of updated states,
is the {\em quantum trajectory} associated to the repeated measurement process.

\noindent
By taking the limit of continuous time,
we arrive at the modern models of continuous observation:
quantum trajectories in continuous time satisfying {\em stochastic
Schr\"odinger equations} \cite{Davies}, \cite{Gisin}, \cite{Car}, \cite{SSE}.
These models are employed with great success for calculations
and computer simulations
of laboratory experiments such as photon counting and homodyne field
detection.


\noindent
In this paper we consider the question,
what happens to the quantum trajectory at large times.
We do so only for the case of discrete time,
not a serious restriction indeed,
since asymptotic behaviour remains basically unaltered
in the continuous time limit.

\noindent
We focus on the case of {\it perfect measurement},
i.e. the situation where no information flows into the system,
and all information which leaks out is indeed observed.
In classical probability such repeated perfect measurement
would lead to a further and further narrowing of the distribution
of the system,
until it either becomes pure, i.e. an atomic measure,
or it remains spread out over some area,
thus leaving a certain amount of information `in the dark' forever.
Using a fundamental inequality of Nielsen \cite{Nielsen}
we prove that in quantum mechanics the situation is quite comparable:
the density matrix tends to purify,
until it hits upon some family of `dark' subspaces, if such exist,
i.e. spaces from which no information can leak out.
A crucial difference with the classical case is, however,
that even after all available information has been extracted by observation,
the state continues to move about in a random fashion between 
the `dark' subspaces,
thus continuing to produce `quantum noise'.

\noindent
The structure of this paper is as follows.
In Section \ref{single} we introduce quantum measurement on a finite system,
in particular Kraus measurement.
In Section \ref{repeat} repeated measurement and the quantum trajectory are
introduced,
and in Section \ref{result} we prove our main result.
Some typical examples of dark subspaces are given in Section \ref{dark}.

\end{section}

\begin{section}
{A single measurement}
\label{single}

\smallskip\noindent
Let $\A$ be the algebra of all complex $d\times d$ matrices.
By $\S$ we denote the space of $d\times d$ {\em density matrices},
i.e. positive matrices of trace 1.
We think of $\A$ as the observable algebra of some finite quantum system,
and of $\S$ as the associated state space.

\noindent
A measurement on this quantum system is an operation
which results in the extraction of information
while possibly changing its state.
Before the measurement the system is described by a {\it prior} state
$\theta\in\S$,
and afterwards we obtain a piece of information,
say an outcome $i\in\{1,2,\ldots,k\}$,
and the system reaches some new (or {\it posterior}) state $\theta'_i$:
   $$\th\longrightarrow(i,\theta'_i)\;.$$
Now, a probabilistic theory,
rather than predicting the outcome $i$,
gives a probability distribution $(\pi_1,\pi_2,\ldots,\pi_k)$
on the possible outcomes.
Let
\begin{equation}
T_i:\th\mapsto\pi_i\th_i'\;,\qquad (i=1,\ldots,k)\;.
\label{defTi}
\end{equation}
Then the operations $T_i$, which must be completely positive,
code for the probabilities $\pi_i=\tr(T_i\th)$ of the possible outcomes,
as well as for the posterior states $\th_i'=T_i\th/\tr(T_i\th)$,
conditioned on these outcomes.
The $k$-tuple $(T_1,\ldots,T_k)$ describes the quantum measurement completely.
Its mean effect on the system,
averaged over all possible outcomes, is given by the trace-preserving map
   $$T:\theta\mapsto\sum_{i=1}^k \pi_i\theta'_i
     =\sum_{i=1}^k T_i\th\;.$$

\smallskip\noindent{\bf Example 1:}
{\sl von Neumann measurement.}

\noindent
Let $p_1,p_2,\ldots p_k$ be mutually orthogonal projections in $\A$
adding up to $\one$,
and let $a\in\A$ be a self-adjoint matrix 
whose eigenspaces are the ranges of the $p_i$.
Then according to von Neumann's projection postulate
a measurement of $a$ is obtained by choosing for $T_i$ the operation
   $$T_i(\theta)=p_i\theta p_i\;.$$

\smallskip\noindent{\bf Example 2:}
{\sl Kraus measurement.}

\noindent
The following indirect measurement procedure
was introduced by Karl Kraus \cite{Kraus}.
It is contains von Neumann's measurement as an ingredient,
but is considerably more flexible and realistic.

\noindent
Our quantum system $\A$ in the state $\theta$ is brought into contact
with a second system, called the `ancilla',
which is described by a matrix algebra $\B$ in the state $\beta$.
The two systems interact for a while under Schr\"odinger's evolution,
which results in a rotation over a unitary $u\in\B\ten\A$.
Then the ancilla is decoupled again,
and is subjected to a von Neumann measurement given by the
orthogonal projections $p_1,\ldots,p_k\in\B$.
The outcome of this measurent contains information about the system,
since system and ancilla have become correlated during their interaction.
In order to assess this information,
let us consider an event in our quantum system,
described by a projection $q\in\A$.
Since each of the projections $p_i\ten\one$ commutes with $\one\ten q$,
the events of seeing outcome $i$ and then the occurrence of $q$
are compatible,
so according to von Neumann we may express the probability for both
of them to happen as:
   $$\PP[\hbox{outcome $i$ and then event $q$}]
  =
\tr\ten\tr\Bigl(\bigl(u(\beta\ten\theta)u^*\bigr)(p_i\ten q)\Bigr)
\;.$$
Therefore the following conditional probability
makes physical sense.
   $$\PP[\hbox{event $q$}|\hbox{outcome $i$}]
={{\tr\ten\tr\Bigl(\bigl(u(\beta\ten\theta)u^*\bigr)(p_i\ten q)\Bigr)}\over
 {\tr\ten\tr\Bigl(\bigl(u(\beta\ten\theta)u^*\bigr)(p_i\ten \one)\Bigr)}}\;.$$
This expression,
which describes the posterior probability of any event $q\in\A$,
can be considered as the posterior state of our quantum system,
conditioned on the measurement of an outcome $i$ on the ancilla,
even when no event $q$ is subsequently measured.
As above, let us therefore call this state $\theta_i'$.
We then have
 $$\tr(\theta_i' q)={{\tr\bigl((T_i \theta)q\bigr)}\over{\tr(T_i\theta)}}\;,$$
where $T_i\th$ takes the form
   $$T_i\theta
   =\tr\ten\id\Bigl(\bigl(u(\beta\ten\theta)u^*\bigr)(p_i\ten\one)\Bigr)\;.$$
Here, id denotes the identity map $\S\to\S$.

\noindent
The expression for $T_i$ takes a simple form
in the case which will interest us here,
namely when the following three conditions are satisfied:

\begin{itemize}
\item[(i)]
$\B$ consists of all $k\times k$-matrices for some $k$;

\item[(ii)]
the orthogonal projections $p_i\in\B$ are one-dimensional
(say $p_i$ is the matrix with $i$-th diagonal entry 1,
and all other entries 0);

\item[(iii)]
$\beta$ is a pure state
(say with state vector $(\beta_1,\ldots,\beta_k)\in\CC^k$).
\end{itemize}

\noindent
These conditions have the following physical interpretations.

\begin{itemize}
\item[(i)]
The ancilla is purely quantummechanical;

\item[(ii)]
the measurement discriminates maximally;

\item[(iii)]
no new information is fed into the system.
\end{itemize}

\noindent
If these conditions are satisfied,
$u$ can be written as a $k\times k$ matrix $(u_{ij})$ of $d\times d$ matrices,
and $T_i$ may be written
\begin{equation}
   T_i\theta=a_i\theta a_i^*\;,
\label{perfectTi}
\end{equation}
where
   $$a_i=\sum_{j=1}^k \beta_j u_{ij}\;.$$
We note that, by construction,
   $$\sok i a_i^*a_i=\sok i\sok j\sok{j'}
     \bar{\beta_j}u_{ji}^*u_{ij'}\beta_{j'}=\|\beta\|^2=\one\;.$$
This basic rule expresses the preservation of the trace by $T$.

\begin{definition}
{\sl By a {\rm perfect measurement} on $\A$ we shall mean a $k$-tuple
$(T_1,\ldots,T_k)$ of operations on $\S$,
where $T_i\theta$ is of the form $a_i\theta a_i^*$ with
$\sok i a_i^*a_i=\one$.}
\end{definition}

\noindent
Mathematically speaking, the measurement $(T_1,\ldots,T_k)$ is perfect iff the 
Stinespring decomposition of each $T_i$ consists of a single term.

\noindent
We note that every perfect Kraus measurement is a perfect measurement
in the above sense,
and that every perfect measurement can be obtained as the result
of a perfect Kraus measurement.
\end{section}

\begin{section}{Repeated measurement}\label{repeat}
By repeating a measurement on the quantum system $\A$ indefinitely,
we obtain a Markov chain with values in the state space $\S$.
This is the quantum trajectory which we study in this paper.

\noindent
Let $\O$ be the space of infinite outcome sequences
$\o=\{\o_1,\o_2,\o_3,\ldots\}$, with $\o_j\in\{1,\ldots,k\}$,
and let for $m\in\NN$ and $\tuple i m\in\{1,\ldots,k\}$ the {\em cylinder
set} $\Lambda_{\tuple i m}\subset\O$ be given by
   $$\Lambda_{\tuple i m}:=\set{\o\in\O}{\o_1=i_1,\ldots,\o_m=i_m}\;.$$
Denote by $\Sigma_m$ the Boolean algebra generated by these cylinder sets,
and by $\Sigma$ the $\sigma$-algebra generated by all these $\Sigma_m$.
Let $T_1,\ldots,T_k$ be as in Section 2.

\noindent
Then for every initial state $\th_0$ on $\A$ there exists a unique probability
measure $\PP_{\th_0}$ on $(\O,\Sigma)$ satisfying
   $$\PP_{\th_0}(\Lambda_{\tuple i m})=\tr(\comprod T i m(\th_0))\;.$$

\noindent
Indeed,
according to the Kolmogorov-Daniell reconstruction theorem we only
need to check consistency: since $T=\sum_{i=1}^k T_i$ preserves the trace,
\begin{eqnarray*}
          \sok i\PP_{\th_0}(\Lambda_{\tuple i m,i})
        &=&\sok i\tr\bigl(T_i\circ\comprod T i m(\th_0)\bigr)
         =\tr(T\circ\comprod T i m(\th_0))\\
        &=&\tr\bigl(\comprod T i m(\th_0)\bigr)
         =\Pthn\bigl(\Lambda_{\tuple i m}\bigr)\;.\\
\end{eqnarray*}

\noindent
On the probability space $(\O,\Sigma,\Pthn)$ we now define the
{\em quantum trajectory} $(\Th_n)_{n\in\NN}$ as the sequence of random
variables given by
   $$\Th_n:\O\to\S:
     \o\mapsto{{\comprod T \o m(\th_0)}
                \over{\tr\bigl(\comprod T \o m(\th_0)\bigr)}}\;.$$
We note that $\Th_n$ is $\Sigma_n$-measurable.
The density matrix $\Th_n(\o)$ describes the state of the system
at time $n$ under the condition that the outcomes
$\o_1,\ldots,\o_n$ have been seen.

\noindent
The quantum trajectory $(\Th_n)_{n\in\NN}$ is a Markov chain with
transitions
\begin{equation}
   \th\longrightarrow\th_i'={{T_i\th}\over{\tr(T_i\th)}}
   \qquad\hbox{with probability}\quad\tr(T_i\th)\;.
   \label{trans}
\end{equation}

\end{section}

\begin{section}{Purification}\label{result}
In a perfect measurement, when $T_i$ is of the form
$\th\mapsto a_i\th a_i^*$,
a pure prior state $\th=\pure\psi$ leads to a pure posterior state:
   $$\th_i'={{a_i\pure{\psi}a_i^*}\over{\inp{\psi}{a_i^*a_i\psi}}}
           =\pure{\psi_i}\;,
\quad\hbox{where}\quad
     \psi_i={{a_i\psi}\over{\|a_i\psi\|}}\;.$$
Hence in the above Markov chain the pure states form a closed set.
Experience with quantum trajectories leads one to believe that in many cases
even more is true:
along  a typical trajectory the density matrix tends to purify:
its spectrum approaches the set $\{0,1\}$.
In Markov chain jargon:
the pure states form an asymptotically stable set.

\noindent
There is, however, an obvious counterexample to this statement in general.
If every $a_i$ is proportional to a unitary,
say $a_i=\sqrt{\l_i}u_i$ with $u_i^*u_i=\one$, then
   $$\th_i'={{a_i\th a_i^*}\over{\tr(a_i\th a_i^*)}}=u_i\th u_i^*\sim\th\;,$$
where $\sim$ denotes unitary equivalence.
So in this case the eigenvalues of the density matrix remain unchanged
along the trajectory:
pure states remain pure
and mixed states remain mixed with unchanging weights.
In this section we shall show that in dimension 2 this is actually
the only exception. (Cf. Corollary \ref{dim2}.)
In higher dimensions the situation is more complicated:
if the state does not purify,
the $a_i$ must be proportional to unitaries on a certain
collection of `dark' subspaces,
which they must map into each other. (Cf. Corollary \ref{spaces}.)

\noindent
In order to study purification we shall consider the {\it moments}
of $\Th_n$.
By the $m$-th {\it moment} of a density matrix $\th\in\S$ we mean
$\tr\left(\th^m\right)$.
We note that two states $\th$ and $\r$ are unitarily equivalent iff
all their moments are equal.
In dimension $d$ equality of the moments $m=1,\ldots,d$ suffices.

\begin{definition}
We say that the quantum trajectory $\bigl(\Th_n(\o)\bigr)_{n\in\NN}$
{\em purifies} when
   $$\forall_{m\in\NN}:\quad\li n\tr\bigl(\Th_n(\o)^m\bigr)=1\;.$$
\end{definition}

\smallskip\noindent
Note that the only density matrices $\r$ satisfying $\tr(\r^m)=1$
are one-dimensional projections, the density matrices of pure states.
In fact, it suffices that the second moment be equal to 1.

\smallskip\noindent
We now state our main result concerning repeated perfect measurement.

\begin{theorem}
Let $\bigl(\Th_n\bigr)_{n\in\NN}$ be the Markov chain with initial
state $\th_0$ and transition probabilities (\ref{trans}).
Then one of the following alternatives holds.

\begin{itemize}

\item[(i)]
The paths of $(\Th_n)_{n\in\NN}$ (the quantum trajectories)
purify with probability 1, or:

\item[(ii)]
there exists a projection $p\in\A$ of dimension at least two
such that
  $$\forall_{i\in\{1,\ldots,k\}}\exists_{\l_i\ge0}:\quad
    pa_i^*a_ip=\l_i p\;.$$

\end{itemize}
\label{main}
\end{theorem}

\noindent
Condition (ii) says that $a_i$ is proportional to an isometry in
restriction to the range of $p$.
Note that this condition trivially holds if $p$ is one-dimensional.

\begin{corollary}
In dimension $d=2$ the quantum trajectory of a repeated perfect measurement
either purifies with probability 1,
or all the $a_i$'s are proportional to unitaries.
\label{dim2}
\end{corollary}

\noindent
If the $a_i$ are all proportional to unitaries,
the coupling to the environment is {\em essentially commutative}
in the sense of \cite{Esscomdil}.

\noindent
Our proof starts from an inequality of Michael Nielsen \cite{Nielsen}
to the effect that for all $m\in\NN$ and all states $\th$:
   $$\sok i\pi_i\,\tr\bigl((\th_i')^m\bigr)\ge \tr(\th^m)\;,$$
where
   $$\pi_i:=\tr(a_i\th a_i^*)
     \quad\hbox{and}\quad 
     \th_i':={{a_i\th a_i^*}\over{\tr(a_i\th a_i^*)}}\;.$$
Nielsen's inequality says that the expected $m$-th moment of the posterior
state is as least as large as the $m$-th moment of the prior state.
In terms of the associated Markov chain we may express this inequality as
   $$\forall_{m,n\in\NN}:\qquad
      \EE\biggl(\tr\left(\Th_{n+1}^m\right)\bigg|\Sigma_n\biggr)
      \ge\tr\left(\Th_n^m\right)\;,$$
i.e. the moments $M_n^{(m)}:=\tr\left(\Th_n^m\right)_{n\in\NN}$
are submartingales.
Clearly all moments take values in $[0,1]$.
Therefore, by the martingale convergence theorem they must
converge almost surely to some random variables $M^{(m)}$.

\noindent
This suggests the following line of proof for our theorem:
Since the moments converge,
the eigenvalues of $(\Th_n)_{n\in\NN}$ must converge.
Hence along a single trajectory
the states eventually become unitarily equivalent,
i.e. eventually
   $$\forall_i:\quad
  \Th_n(\o)\sim{{a_i\Th_n(\o)a_i^*}\over{\tr(a_i\Th_n(\o)a_i^*)}}\;.$$
But this seems to imply that either $\Th_n$ purifies almost surely,
or the $a_i$'s are unitary on the support of $\Th_n$.

\noindent
In the following proof of Theorem \ref{main} we shall make this suggestion
mathematically precise.

\begin{lemma}
In the situation of Theorem \ref{main}
one of the following alternatives holds.

\begin{itemize}

\item[(i)]
For all $m\in\NN$: $\displaystyle\li n\tr\left(\Th_n^m\right)=1$ almost surely;

\item[(ii)]
there exists a mixed state $\r\in\S$ such that
   $$\forall_{i=1,\ldots,k}\exists_{\l_i\ge0}:\quad
     a_i\r a_i^*\sim\l_i\r\;.$$

\end{itemize}
\label{stochlemma}
\end{lemma}

\smallskip\noindent{\it Proof.}
For each $m\in\NN$ we consider the continuous function
   $$\d_m:\S\to[0,\infty):\th\mapsto\sok i\tr(a_i\th a_i^*)
 \left(\tr\left(\left({{a_i\th a_i^*}\over{\tr(a_i\th a_i^*)}}\right)^m\right)
           -\tr(\th^m)\right)^2\;.$$
Then, using (\ref{perfectTi}) and (\ref{trans}),
   $$\d_m(\Th_n)
   =\EE\biggl(\left(M_{n+1}
^{(m)}-M_n^{(m)}\right)^2\bigg|\Sigma_n\biggr)\;.$$
Since $\bigl(M_n^{(m)}\bigr)_{n\in\NN}$ is a positive submartingale
bounded by 1, its increments must be square summable:
   $$\forall_{m\in\NN}:\qquad
     \sum_{n=0}^\infty\EE\bigl(\d_m(\Th_n)\bigr)\le1\;.$$
In particular

\begin{equation}
\li n\sod m\EE\bigl(\d_m(\Th_n)\bigr)=0\;.
\label{deltatozero}
\end{equation}

\noindent
Now let us assume that $(i)$ is not the case,
i.e. for some (and hence for all) $m\ge2$ the expectation
$\EE(M^{(m)})=:\mu_m$ is strictly less than 1.
For any $n\in\NN$ consider the event
   $$A_n:=\Set{\o\in\O}{M_n^{(2)}\le{{\mu_2+1}\over2}}\;.$$
Then,
since $\EE\left(M_n^{(2)}\right)$ is increasing in $n$,
we have for all $n\in\NN$:
   $$\EE(M_n^{(2)})\le\EE(M^{(2)})=\mu_2<1\;.$$
Therefore for all $n\in\NN$,

\begin{eqnarray*}
\mu_2&\ge&\EE\left(M_n^{(2)}\cdot1_{[M_n^{(2)}>{{\mu_2+1}\over2}]}\right)\\
     &\ge&{{\mu_2+1}\over2}\;\PP\left[M_n^{(2)}>{{\mu_2+1}\over2}\right]\\
       &=&{{\mu_2+1}\over2}\left(1-\PP(A_n)\right)\;,
\end{eqnarray*}
so that
\begin{equation}
   \PP(A_n)\ge{{1-\mu_2}\over{1+\mu_2}}\;.
\label{lowboundPAn}
\end{equation}
On the other hand,
$A_n$ is $\Sigma_n$-measureable and therefore it is a union of sets
of the from $\Lambda_{\tuple i n}$.
Since $\Th_n$ is $\Sigma_n$-measureable,
$\Th_n$ is constant on such sets; let us call the constant
$\Th_n(\tuple i n)$.
We have the following inequality:
   $${1\over{\PP(A_n)}}\sum_{\Lin\subset A_n}\PP\left(\Lin\right)
     \left(\sod m\d_m\bigl(\Th_n(\tuple i n)\bigr)\right)
     \le {1\over{\PP(A_n)}}\sod m \EE\bigl(\d_m(\Th_n)\bigr)\;.$$
On the left hand side we have an average of numbers
which are each of the form
$\sod m\d_m(\Th_n(\tuple i n))$,
hence we can choose $(\tuple i n)$ such that $\r_n:=\Th_n(\tuple i n)$
satisfies, by (\ref{lowboundPAn}),
   $$\sod m\d_m(\r_n)\le
   {{\mu_2+1}\over{\mu_2-1}}\;\sod m\EE\bigl(\d_m(\Th_n)\bigr)\;.$$
Since $\Lin\subset A_n$,
the sequence $(\r_n)_{n\in\NN}$ lies entirely in the compact set
   $$\Set{\th\in\S}{\tr(\th^2)\le{{\mu_2+1}\over2}}\;.$$
Let $\r$ be a cluster point of this sequence.
Then, since $\EE(\d_m(\Th_n))$ tends to 0 as $n\to\infty$,
and $\d_m$ is continuous, we may conclude that for $m=1,\ldots,d$:
   $$\d_m(\r)=0\;,\quad\hbox{and}\quad\tr(\r^2)\le{{\mu_2+1}\over2}<1\;.$$
So $\r$ is a mixed state, and,
by the definition of $\d_m$,
   $$\tr(a_i\r a_i^*)
\left(\tr\left(\left({{a_i\r a_i^*}\over{\tr(a_i\r a_i^*)}}\right)^m\right)
           -\tr(\r^m)\right)^2=0$$
for all $m=1,2,3,\ldots,d$ and all $i=1,\ldots,k$.
Therefore either $\tr(a_i\r a_i^*)=0$,
i.e. $a_i\r a_i^*=0$,
proving our statement $(ii)$ with $\l_i=0$;
or $\tr(a_i\r a_i^*)>0$,
in which case $\r_i':=a_i\r a_i^*/\tr(a_i\r a_i^*)$ and $\r$ itself have
the same moments of orders $m=1,2,\ldots,d$,
so that they are unitarily equivalent.
This proves $(ii)$.

\qed

\noindent
From Lemma \ref{stochlemma} to Theorem \ref{main}
is an exercise in linear algebra:

\begin{lemma}
Let $a_1,\ldots,a_k\in M_d$ be such that $\sok i a_i^*a_i=\one$.
Suppose that there exists a density matrix $\r\in M_d$ such that
for $i=1,\ldots,k$
   $$a_i\r a_i^*\sim \l_i\r\;.$$
Let $p$ denote the support of $\r$. Then for all $i=1,\ldots,k$:
   $$pa_i^*a_i p=\l_i p\;.$$
\label{linalglemma}
\end{lemma}

\smallskip\noindent{\it Proof.}
Let us define, for a nonnegative matrix $x$,
the {\it positive determinant} $\detpos(x)$
to be the product of all its strictly positive eigenvalues
(counted with their multiplicities).
Then, if $p$ denotes the support projection of $x$, we have the implication
\begin{equation}
\detpos(x)=\detpos(\l p)\quad\Longrightarrow\quad\tr(xp)\ge\tr(\l p)
\label{detposineq}
\end{equation}
with equality iff $x=\l p$.
(This follows from the fact that the sum of a set of positive numbers with given
product is minimal iff these numbers are equal.) 

\noindent
Now let $p$ be the support of $\r$ as in the Lemma.
Let $v_i\sqrt{p a_i^*a_i p}$
denote the polar decomposition of $a_i p$.
Then we have by assumption,

\begin{eqnarray*}
   \detpos(\l_i\r)&=&\detpos(a_i\r a_i^*)\\
                  &=&\detpos(a_i p\r p a_i^*)\\
            &=&\detpos(v_i\sqrt{p a_i^*a_i p}\r\sqrt{p a_i^*a_i p}v_i^*)\\
            &=&\detpos(\sqrt{p a_i^*a_i p}\r\sqrt{p a_i^*a_i p})\\
            &=&\detpos(pa_i^*a_ip)\detpos(\r)\;.
\end{eqnarray*}
Now, since $\detpos(\l_i\r)=\detpos(\l_i p)\cdot\detpos(\r)$ and $\detpos(\r)>0$,
it follows that
   $$\detpos(\l_i p)=\detpos(pa_i^*a_ip)\;.$$
By the implication (\ref{detposineq}) we may conclude that
\begin{equation}
\tr(pa_i^*a_ip)\ge\tr\l_i p)\;.
\label{paapineq}
\end{equation}
On the other hand,

 $$\sok i \l_i= \sok i \tr(\l_i \r)\\
              = \sok i \tr(a_i\r a_i^*)
              = \tr\left(\r\left(\sok i a_i a_i^*\right)\right)
              = \tr\r=1\;,$$
where in the second equality sign the assumption was used again.
Then, by (\ref{paapineq}),
   $$\tr\,p  = \sok i \tr(pa_i^*a_ip) \ge \sok i \tr(\l_i p)
            = \left(\sok i \l_i\right)\tr p = \tr\,p\;.$$
So apparently, in this chain, we have equality.
But then, since equality is reached in (\ref{detposineq}),
we find that
   $$pa_i^*a_ip = \l_i p\;.$$
\qed

\end{section}

\begin{section}{Dark subspaces}\label{dark}
By considering more than one step at a time the following stronger conclusion
can be drawn.

\begin{corollary}
In the situation of Theorem \ref{main},
either the quantum trajectory purifies with probability 1 or there exists a projection
$p$ of dimension at least 2 such that for all $l\in\NN$ and all $\tuple i l$
there is $\l_{\tuple i l}\ge0$ with
\begin{equation}
pa_{i_1}^*\cdots a_{i_l}^*a_{i_l}\cdots a_{i_1}p=\l_{\tuple i l}p\;.
\label{spaces}
\end{equation}
\end{corollary}

\noindent
We shall call a projection $p$ satisfying (\ref{spaces})
a {\em dark} projection, and its range a {\em dark} subspace.

\noindent
Let $p$ be a dark projection,
and let $v_i\sqrt{pa_i^*a_i}=\sqrt{\l_i}v_ip$
be the polar decomposition of $a_ip$.
Then the projection $p_i':=v_ipv_i^*$ satisfies:

\begin{eqnarray*}
         \l_i p_i' a_{i_1}^*\cdots a_{i_m}^* a_{i_m}\cdots a_{i_1} p_i'
      &=&\l_i (v_i p v_i^*)a_{i_1}^*\cdots a_{i_m}^* 
             a_{i_m}\cdots a_{i_1} (v_i p v_i^*)\\
      &=&v_i p a_i^*a_{i_1}^*\cdots a_{i_m}^* 
             a_{i_m}\cdots a_{i_1} a_i p v_i^*\\
      &=&\l_{i,\tuple i m}\cdot p_i'\;.
\end{eqnarray*}
Hence if $p$ is dark, and $\l_i\ne0$
then also $p_i'$ is dark with constants
   $$\l'_{\tuple i m}=\l_{i,\tuple i m}/\l_i\;.$$

\noindent
We conclude that asymptotically the quantum trajectory performs a random walk
between dark subspaces of the same dimension,
with transition probabilities $p\longrightarrow p_i'$ equal to
$\l_i$, the scalar value in $pa_i^*a_ip=\l_ip$.
In the trivial case that the dimension of $p$ is 1,
purification has occurred.

\noindent
Inspection of the $a_i$ should reveal the existence of nontrivial dark
subspaces.
If none exist, then purification is certain.

\noindent
We end this Section with two examples where nontrivial dark subspaces occur.

\smallskip\noindent{\bf Example 1.}
Let $d=l\cdot e$ and let $\H_1,\ldots,\H_l$ be
mutually orthogonal $e$-dimensional subspaces of $\H=\CC^d$.
Let $(\pi_{ij})$ be an $l\times l$ matrix of transition probabilities.
Define $a_{ij}\in\A$ by
   $$a_{ij}:=\sqrt{\pi_{ij}}v_{ij}\;,$$
where the maps $v_{ij}:\H_i\to\H_j$ are isometric.
Then the matrices $a_{ij}$, $i,j=1,\ldots,l$ define a perfect measurement
whose dark subspaces are $\H_1,\ldots,\H_l$.

\goodbreak
\smallskip\noindent{\bf Example 2.}

\nobreak\noindent
The following example makes clear that nontrivial dark subspaces need not be
orthogonal.

\noindent
Let $\H=\CC^2\otimes\D$,
where $\D$ is some finite dimensional Hilbert space,
and for $i=1,\ldots,k$ let $a_i:=b_i\otimes u_i$,
where the $2\times 2$-matrices $b_i$ satisfy the usual equality
   $$\sok i b_i^*b_i=\one\;,$$
and the $u_i$ are unitaries $\D\to\D$.
Suppose that the $b_i$ are not all proportional to unitaries.
Then the quantum trajectory defined by the $a_i$ has dark subspaces
$\psi\otimes\D$, with $\psi$ running through the unit vectors in $\CC^2$.
Physically this example describes a pair of systems without any interaction
between them,
one of which is coupled to the environment in an
essentially commutative way, whereas the other purifies.

\end{section}

\end{document}